\documentstyle[12pt]{article}
\begin{document}

\def\beq{\begin{equation}}
\def\eeq{\end{equation}}
\def\bce{\begin{center}}
\def\ece{\end{center}}
\def\bea{\begin{eqnarray}}
\def\eea{\end{eqnarray}}
\def\ben{\begin{enumerate}}
\def\een{\end{enumerate}}
\def\ul{\underline}
\def\ni{\noindent}
\def\nn{\nonumber}
\def\bs{\bigskip}
\def\ms{\medskip}
\def\wt{\widetilde}
\def\wh{\widehat}
\def\brr{\begin{array}}
\def\err{\end{array}}
\newcommand{\sm}[1]{\sum_{#1 =0}^{\infty}}
\newcommand{\Gm}{\Gamma}
\newcommand{\ld}{\lambda}
\newcommand{\itg}{\int_0^\infty}
\newcommand{\nud}{\frac{d+1}{2}}
\newcommand{\sma}[1]{\sum_{j=0}^{d-1} e_{j}(d,#1)}
\newcommand{\nd}{\frac{d+1}{2}}
\newcommand{\del}[1]{\frac{\partial}{\partial #1}}
\newcommand{\oB}{\vert_{\partial M}=0}

\hfill UB-ECM-PF 96/7

\hfill March, 1996


\bce
{\bf \Large Zeta function for the Laplace operator acting on \\
 forms in a ball with gauge boundary conditions}
\ece
\vspace{5mm}
\bce
{\bf E. Elizalde}\footnote{E-mail: eli@zeta.ecm.ub.es},
{\bf M. Lygren}\footnote{E-mail: lygren@fm.unit.no}\\
Center for Advanced Studies CEAB, CSIC, Cam\'{\i} de Santa
B\`arbara,
17300 Blanes,
\\ and  Department ECM and IFAE, Faculty of Physics,
University of Barcelona, \\ Diagonal 647, 08028 Barcelona,
Spain \\
and \\
{\bf D.V. Vassilevich} \\
Department of Theoretical Physics, St. Petersburg University, \\
198904 St. Petersburg, Russia
\ece
\begin{abstract}
The Laplace operator acting on antisymmetric
tensor fields in a $D$--dimensional Euclidean ball
is studied.
Gauge-invariant local boundary conditions
(absolute and relative ones, in the language of Gilkey) are
considered. The eigenfuctions of the operator are found explicitly for
all values of $D$. Using in a row a number of basic techniques, as
Mellin transforms, deformation and shifting of the complex
integration contour,
and pole compensation,  the zeta function of the operator is
obtained. From its expression, in particular,
 $\zeta (0)$ and $\zeta'(0)$ are evaluated exactly. A table is given
in the paper  for $D=3, 4, ...,8$.
The functional determinants and Casimir energies are
obtained  for $D=3, 4, ...,6$.
\end{abstract}

\vfill

\newpage

\section{Introduction}
In this paper we obtain the zeta function of the Laplace operator acting
on antisymmetric tensor fields defined in a $D$-dimensional ball with
gauge-invariant boundary conditions. Mathematically this computation is
quite an imposing challenge, as is proven by the number of erroneous
results reported in the literature on this and related computations
(details will be given later).
The physical motivations for such a study are to be
found in quantum cosmology, where the $\zeta$ function of the Laplacian
describes the contribution of antisymmetric tensor fields
and ghosts to the pre-factor of the wave function of the
universe (see e.g. Ref. \cite{Esp}). An intriguing problem in this
context is the non-compensation of the boundary contributions to the
one--loop divergences between different members of the
supergravity supermultiplet \cite{Mp}.
Another motivation is
to provide the numerical material needed to extend previous
analysis of the heat kernel asymptotics \cite{BGV}
to the case of mixed boundary conditions.

There are two admisible sets
of gauge--invariant local boundary conditions ---which
have been called by Gilkey, respectively, absolute and relative boundary conditions
\cite{PBG}. These sets are dual to each other and are becoming highly
interesting in connection with recent developments in string theory. Hence
one can study $p$-forms with $p\le \left [ \frac {D+1}2
\right ]$ for both types of boundary conditions. Due to
duality, the determinant of the Laplacian for $p$-forms with
absolute boundary conditions is the same as for $(D-p)$--forms
with relative boundary conditions. Furthermore, owing to
gauge--invariance we can restrict ourselves to transversal
$p$-forms. The complete result will just be a sum of the ones
for transversal $p$-- and $(p-1)$--forms, provided zero modes
are properly taken into account. To find the spectrum of
the Laplace operator on transversal $p$-forms we use the
 powerful
method proposed in Refs. \cite{vass2,Dplb}.
They involve
integral representations of the spectral sum, Mellin transformations,
non-trivial commutation of series and integrals and skillful
analytic continuation of zeta functions on the complex plane.
Here we will focus again on a class of situations
 for which the eigenvalues of the differential operator, $A$,
 are not known explicitly, but where nevertheless the exact calculation
of $\ln\det A$ is possible. The method  is applicable whenever an
implicit
equation satisfied by the eigenvalues is at hand and some asymptoticity
 properties  of the equation are known too.

In this paper, we will find explicit
solutions of the $D$-dimensional transversality condition
in terms of $p$-- and $(p-1)$--forms obeying a $(D-1)$--dimensional
transversality condition. These forms will satisfy now pure
Dirichlet or Robin boundary conditions, instead of mixed
absolute or relative boundary conditions. This clever procedure
will enable us
to find exact eigenfunctions and to express the eigenvalues of
the Laplace operator in terms of Bessel functions and
their derivatives. After that, we will be able to perform explicitly the
evaluation of the zeta function at the origin, and of the
determinant of the Laplacian as well. We will also calculate the Casimir
energy. A table of results is given in the paper
for $D=3,4,5,6,7,8$, which cover the situations which appear in the
usual supersymmetric
theories. However, our expressions to come are actually valid (and can
be used) for any dimension $D$ and yield explicit, exact values in a
reasonable
amount of algebraic computation time. Usual methods for the acceleration
of the series convergence improve performance considerably.

In connection with previous results, we should point out that for some
scattered values of $D=4$ and $p=1,2$, several first heat-kernel
coefficients have been calculated in the papers
\cite{vass2}-\cite{EKMP}. These results agree with the analytical
formulas in \cite{BG} once the corrections that were found in Ref.
 \cite{vass2} are taken
into account (see also Ref. \cite{Mp}). For $D=4,\ p=1$, the
one--loop effective action has been evaluated in \cite{EKP} for
a specific choice of gauge and boundary conditions.

The paper is organized as follows. In Sect. 2 we use the Hodge-de Rham
decomposition of $p$-forms in order to simplify the structure of the
 spectrum of the Laplacian  operator in the $D$-dimensional ball and,
subsequently, of the corresponding determinant.
After writing the absolute and
relative boundary conditions as Dirichlet and Robin ones, a
convenient analytical
continuation of the associated zeta function is performed in Sect. 3, in
some detail, what leaves us in a position wherefrom we can find the
heat-kernel coefficients, the determinant, Casimir energies, and so on.
The calculation of the zeta function at the origin is undertaken in Sect.
4, that of the determinant in Sect. 5, and the Casimir energy is obtained
in Sect. 6. Finally, in an Appendix we give an exhaustive
 list of all the determinants
obtained, both for the case of absolute and
relative boundary conditions.
\bs

\section{Spectrum of the Laplace operator in a ball}

Consider the $D=d+1$ dimensional unit disk with the metric
\begin{equation}
ds^2=dr^2+r^2 d\Omega^2, \quad 0 \le r \le 1, \label{eq:(2)}
\end{equation}
where $d\Omega^2$ is the metric on the unit sphere $S^d$.
Throughout this paper we shall use the notations $\{ x_\mu \} $$=
\{ x_0,x_i\}$, $x^0=r$, $\mu =0,1,...,d$. The $(d+1)$--dimensional
Laplace operator $\Delta =\nabla^\mu
\nabla_\mu$ acting on a $p$-form, $B$, can be written as
\begin{eqnarray}
(\Delta B)_{i_1...i_{p-1}0}= \left ( \partial_0^2+
\frac {d-2p+2}r \partial_0 +\frac {p^2-dp-1}{r^2}
+\ ^{(d)}\Delta \right ) B_{i_1...i_{p-1}0}-\nonumber \\
-\frac 2r \ ^{(d)}\nabla^k B_{i_1...i_{p-1}k}
\label{eq:(3)} \\
(\Delta B)_{i_1...i_p}= \left ( \partial_0^2+
\frac {d-2p}r \partial_0 +\frac {p^2-dp}{r^2}
+\ ^{(d)}\Delta \right ) B_{i_1...i_p}+ \nonumber \\
+\frac 2r \sum_{a=1}^p \ ^{(d)}\nabla_{i_a}
B_{i_1...i_{a-1}0i_{a+1}...i_p} , \label{eq:(4)}
\end{eqnarray}
where $\ ^{(d)}\nabla$ and $\ ^{(d)}\Delta$ are the
covariant derivative and Laplace operator corresponding
to the $d$-dimensional metric $g_{ik}$.

    Any $p$-form $B^p$ admits the Hodge-de Rham decomposition:
\begin{equation}
B^p=B^{p\perp}+dB^{(p-1)\perp} , \label{eq:hdr}
\end{equation}
where $B^{p\perp}$ denotes a transversal $p$-form. The
decomposition (\ref{eq:hdr}) commutes with the Laplace
operator. Thus, in order to define the spectrum of the Laplacian on
the space of all antisymmetric forms, it is enough to
study the case of transversal forms only.

    There are two sets of local boundary conditions
consistent with the decomposition (\ref{eq:hdr}).
They are the so-called absolute and relative boundary
conditions \cite{PBG}. In the coordinate system (\ref{eq:(2)})
the absolute boundary conditions read
\begin{equation}
\partial_0 B_{i_1,...,i_p}\oB , \quad
B_{0,i_1,...,i_{p-1}}\oB ,\label{eq:abs}
\end{equation}
while the relative boundary conditions have the form
\begin{equation}
B_{i_1,...,i_p}\oB ,\quad
\left (\partial_0+\frac {d-2p+2}r \right )
B_{0,i_1,...,i_{p-1}}\oB . \label{eq:rel}
\end{equation}
Consider the $(d+1)$--dimensional transversality condition
\begin{equation}
\nabla^\mu B_{\mu \nu ...\rho}=0 . \label{eq:(5)}
\end{equation}
On a disk it can be written as
\begin{eqnarray}
(\nabla B)_{i_1...i_{p-2}0}&=&\ ^{(d)}\nabla^i
B_{ii_1...i_{p-2}0}=0,\label{eq:(6)} \\
(\nabla B)_{i_1...i_{p-1}}&=&(\partial_0+
\frac {d-2p+2}r )B_{0i_1...i_{p-1}}+
\ ^{(d)}\nabla^i B_{ii_1...i_{p-1}} =0.\label{eq:(7)}
\end{eqnarray}
According to the general method developed in the papers \cite{vass2,Dplb},
the solutions of the equations (\ref{eq:(6)}) and (\ref{eq:(7)})
can be expressed in terms of $d$--dimensional transversal
forms:
\begin{equation}
B^{p\perp}=B^{pT}+B^{p\perp}(\psi^{(p-1)T}),
\label{eq:sol}
\end{equation}
where such $d$--dimensional transversal forms satisfy the equations:
\begin{equation}
\ ^{(d)}\nabla^i A^{pT}_{i,i_1,...,i_{p-1}}=0, \qquad
A^{pT}_{0,i_1,...,i_{p-1}}=0. \label{eq:defT}
\end{equation}
Here $A^T$ denotes either $B^T$ or $\psi^T$. The second
term in (\ref{eq:sol}) has the form:
\begin{eqnarray}
B^\perp_{0i_1...i_{p-1}}(\psi^T )&=&(-^{(d)}\Delta +
\frac {(p-1)d-(p-1)^2}{r^2})r \psi^T_{i_1...i_{p-1}},
 \nonumber \\
B^\perp_{i_1...i_p}(\psi^T)&=&(\partial_0+\frac {d-2p}r)
r\ ^{(d)}\nabla_{[i_1}
\psi^T_{i_2...i_p]} \label{eq:bpsi} \\
\ & \ &\nabla_{[i_1}\psi_{i_2...i_p]}:=\sum_{n=1}^p
(-1)^{n+1}\nabla_{i_n} \psi_{i_1...i_{n-1}i_{n+1}...i_p}
\nonumber
\end{eqnarray}
One can prove that the Laplace operator (\ref{eq:(3)}),
(\ref{eq:(4)}), commutes with the decomposition (\ref{eq:sol}):
\begin{equation}
\Delta B^{p\perp}=\Delta B^{pT}+
B^{p\perp}(\Delta \psi^{(p-1)T}).
\label{eq:com}
\end{equation}
The determinant of the Laplace operator on the space of
$(d+1)$--dimensional transversal $p$ forms can be represented
as a product of two determinants, taken over $d$-dimensional
transversal $p$-- and $(p-1)$--forms:
\begin{equation}
\det (-\Delta )_{p\perp}=\det (-\Delta )_{pT}
\times \det (-\Delta )_{(p-1)T}. \label{eq:fact}
\end{equation}
Moreover, the fields $B^T$ and $\psi^T$ satisfy pure
boundary conditions. The boundary conditions for
$B^T$ are defined by the first equations in (\ref{eq:abs})
and (\ref{eq:rel}), for absolute and relative boundary
conditions, respectively. For absolute boundary conditions
on the field $B^{p\perp}$, the form $\psi^{T(p-1)}$
satisfies Dirichlet boundary conditions
\begin{equation}
\psi^{T}\oB . \label{eq:psia}
\end{equation}
For relative boundary conditions we have, for the $(p-1)$--form
$\psi^T$,
\beq
\left ( \partial_0 +\frac {d-2p+1}r \right ) \psi^T \oB,
\label{eq:psir}
\eeq
that is, Robin boundary conditions.
We thus see that the initial eigenvalue problem for
the
$(d+1)$--dimensional transversal $p$-forms with mixed
boundary conditions is reduced to two eigenvalue
problems, for $d$-dimensional transversal $p$-- and
$(p-1)$--forms with pure boundary conditions (Dirichlet and Robin).

In the particular cases when $p=1,2$, the boundary conditions
(\ref{eq:psia}) and (\ref{eq:psir}) agree with the corresponding
expressions
of papers \cite{vass2,Dplb}.
Note that the $l=0$ scalar mode generates a zero mode of
the mapping $\psi \to B^{1\perp}$. Hence this mode should
be excluded when one considers the path integral over
transversal 1-forms and from the second determinant
on the r.h.s. of Eq. (\ref{eq:fact}).

Let us introduce the set of $d$-dimensional spherical harmonics,
$Y^{(l)p}_{i_1,...,i_p} (x_j)$, corresponding to transversal
$p$-forms on $S^d$.
They are eigenmodes of the $d$-dimensional
Laplacian. The associated eigenvalues and degeneracies $D_l^p$
are  found to be \cite{IK,elv}
\begin{eqnarray}
&& ^{(d)}\Delta
 Y^{(l)p}_{i_1,...,i_p} (x_j)
\ = \ \frac 1{r^2} [-l(l+d-1)+p]
Y^{(l)p}_{i_1,...,i_p} (x_j), \nn \\
&& D_l^p  \ = \ \frac{(2l+d-1) \ (l+d-1)!}{p! \, (d-p-1)! \, (l-1)! \,
(l+p) (l+d-p-1)} .
\label{eq:lapl}
\end{eqnarray}
We can represent the eigenfunctions of the complete
$(D=d+1)$--dimensional Laplace operator as a Fourier series
in the harmonics (\ref{eq:lapl}):
\begin{equation}
B^{pT}_{i_1,...,i_p}(r,x_j)=\sum_{(l)}
Y^{(l)p}_{i_1,...,i_p} (x_j)
f{(l)}(r).
\label{eq:Four}
\end{equation}
Here we need to sum  over  $(l)$, which means summation
over the index $l$ from 1 to $\infty$ and  over another
index, from 1 to $D_l^p$, which describes the different harmonics
with degenerate eigenvalues of the Laplacian $\ ^{(d)}\Delta$. This last
summation is not shown explicitly.

We can now substitute the decomposition (\ref{eq:Four})
in the eigenvalue equation
\begin{equation}
\Delta B^{pT}_{\mu_1, \mu_2,...,\mu_p}=
-\lambda^2 B^{pT}_{\mu_1,\mu_2,...,\mu_p}
\label{eq:eigen}
\end{equation}
for the $(d+1)$--dimensional Laplacian (\ref{eq:(3)}),
(\ref{eq:(4)}). Let us recall the fact that for the fields
$B^{pT}$ the zero-th components vanish identically:
$B^{pT}_{0i_1 ...i_{p-1}}=0$. The equation for the
 components (\ref{eq:(3)})
 reduces to the trivial
 identity $0=0$. The other components (\ref{eq:(4)})
lead to an equation of Bessel type for $f^{(l)}(r)$.
After a rather lengthy algebra, one finds that the eigenfunctions
of the Laplace operator (\ref{eq:(4)}) have the following
form:
\begin{equation}
r^{(1-d)/2+p} J_{(d-1)/2+l}(\lambda_l r)
Y^{(l)p}_{i_1,...,i_p} (x_j),
\label{eq:eig}
\end{equation}
where $J_n$ denote
Bessel functions. The eigenvalues $\lambda_l^2$ are defined
by boundary conditions and their degeneracies $D_l^p$ are
given by (\ref{eq:lapl}).

>From the preceding expressions, we are able to evaluate  the
determinant of the Laplace
operator on the space of transversal $p$-forms. We obtain
\begin{equation}
{\det}_{p\perp} (-\Delta )=
\prod_{l=1}^\infty \lambda^{2D_l^p}_l
\prod_{k=1}^\infty \kappa^{2D_k^{p-1}}_k,
\label{eq:dete}
\end{equation}
where for absolute boundary conditions the eigenvalues
$\lambda$ and $\kappa$ are defined by (\ref{eq:abs}) and
(\ref{eq:psia}), namely
\begin{equation}
\partial_0
r^{(1-d)/2+p} J_{(d-1)/2+l}(\lambda_l r) \oB ,
\qquad J_{(d-1)/2+k}(\kappa_k r) \oB.
\label{eq:eiga}
\end{equation}
For relative boundary conditions we have, from (\ref{eq:rel})
and (\ref{eq:psir}),
\begin{equation}
J_{(d-1)/2+l} (\lambda_l r) \oB, \qquad
(\partial_0 +d-2p+1)
r^{(1-d)/2+p-1} J_{(d-1)/2+k}(\kappa_k r) \oB .
\label{eq:eigr}
\end{equation}
One can easily check that the eigenfunctions defined
in this section satisfy all necessary orthogonality
properties.
\bs

\section{Analytical continuation of the zeta function}

Both the absolute and the relative boundary conditions can be written as
Dirichlet and Robin boundary conditions, in the form
\beq
\left. J_\nu (\lambda_l r)\right|_{\partial M} =0, \label{dirichlet}
\eeq
\beq
\left. u(d,p) J_\nu (\lambda_l r)+\lambda_l
       J'_\nu (\lambda_l r)\right|_{\partial M}=0. \label{robin}
\eeq
For the absolute boundary conditions we have $u(d,p)=(1-d)/2+p$,
 while for the relative
boundary condition, $u(d,p)=(1+d)/2-p$.
The boundary $\partial M$ is here described by $r=a$.
The zeta function is
\beq
\zeta^{d}_{p T} (s)= \sum_{l=1}^\infty D_l^p(d) \lambda^{-2s}.
\eeq
The decomposition described in the last section will, at the level of
zeta functions, manifest itself as a sum of the different zeta functions
belonging to each term of the decomposition.
Thus we can write
\beq
\zeta^{d}_{p \perp } (s)=\zeta^{d}_{p T} (s)+\zeta^{d}_{(p-1) T} (s).
\eeq
The zeta function is in general convergent for $s>\frac{d+1}{2}$ only,
but it
can be analytically continued in the complex plane to all values of $s$,
in particular to the vicnity of $s=0$. Several authors have considered
zeta
 functions corresponding to operators which eigenvalues are not given
explicitly. In special,
 they have investigated in detail the case
when they are given under the form of roots of equations involving
Bessel functions (see, for example~\cite{bek}).
In the Refs.~\cite{bek,bgke} it has been shown explicitly how this
analytical continuation can be carried out for zeta functions of this
kind, and we will follow
them closely. The reader may resort to those papers for all
particularities skipped from  the present calculation.

Writing the boundary conditions (\ref{dirichlet}) and (\ref{robin})
symbolically as
$\left. \Psi_\nu
  (\lambda_l r)\right|_{\partial M}=0$, the first idea is to express
 the zeta
function as a contour integral along a path $\gamma$
enclosing all positive
solutions of the boundary condition equation, namely
\beq
\zeta^{d}_{p\bot}(s)=\sum_{l=1}^\infty  D_l(p,d) \int_\gamma \frac{dk}{2\pi}
           (k^2+m^2)^{-s} \frac{\partial }{\partial k} \ln  \Psi_\nu (ak).
\eeq
Here we have introduced the constant $m$ in order
 to simplify the treatement of
the problem. It is, however, not essential to obtain the final result
and we shall let this constant vanish later in the calculation. To
be able to use most of the techniques  developed in the
papers~\cite{bek,bgke}, we  expand
the degeneracy as
\beq
 D_l(p,d)=\sma{p} \left(l+\frac{d-1}{2}\right)^{j}.
\eeq
The zeta function  reads then
\bea
\zeta^{d}_{p\bot}(s)&=&\sma{p}\sum_{l=1}^\infty\left(l+\frac{d-1}{2}\right)^{j}
   \int_\gamma \frac{dk}{2\pi}
           (k^2+m^2)^{-s} \frac{\partial }{\partial k} \ln
            \Psi_{l+\frac{d-1}{2}} (ak)  \nn \\
&=&\sma{p}\sum_{l=0}^\infty\left(l+\frac{d+1}{2}\right)^{j}
   \int_\gamma \frac{dk}{2\pi}
           (k^2+m^2)^{-s} \frac{\partial }{\partial k} \ln
            \Psi_{l+\frac{d+1}{2}} (ak).
\eea
Now we are in the position of doing the analytic continuation.
This involves
substracting and adding the leading asymptotic terms of the
uniform expansion of the Bessel function
$I_\nu(k)$ and its derivative. For  $\nu \rightarrow \infty$
and $z=k/\nu$ being fixed, they are~\cite{abram}:
\beq
I_\nu(\nu z) \sim \frac{1}{\sqrt{2\pi \nu}}\frac{e^{\nu\mu}}{(1+z^2)^{1/4}}
       \left[ 1+ \sum_{k=1}^\infty \frac{u_k(t)}{\nu^k} \right]
\eeq
and
\beq
I'_\nu(\nu z) \sim \frac{1}{\sqrt{2\pi \nu}} \frac{e^{\nu\mu}(1+z^2)^{1/4}}{z}
       \left[ 1+ \sum_{k=1}^\infty \frac{v_k(t)}{\nu^k} \right],
\eeq
respectively. Here
$u_k$ and $v_k$ are functions obtained in a recursive way
  in~\cite{abram}, while
$t=1/\sqrt{1+z^2}$ and $\mu=\sqrt{1+z^2}+\ln[z/(1+ \sqrt{1+z^2})]$.
Furthermore, we define the coefficients $D_n(t)$ and $M_n(t)$ by
\beq
\ln  \left[ 1+ \sum_{k=1}^\infty \frac{u_k(t)}{\nu^k} \right] \sim
  \sum_{n=1}^\infty \frac{D_n(t)}{\nu^n}
\eeq
and
\beq
\ln  \left[ 1+ \sum_{k=1}^\infty \frac{v_k(t)}{\nu^k}+
   \frac{u(p,d)}{\nu}t \left(1+\sum_{k=1}^\infty \frac{u_k(t)}{\nu^k} \right)
    \right] \sim
    \sum_{n=1}^\infty \frac{M_n(p,d)(t)}{\nu^n}.
\eeq
Then, by adding and subtracting the first $N$ terms of these last two
expansions, we can write the zeta function for Dirichlet boundary
conditions as
\beq
\zeta^{d}_{p\bot}(s)=\sum_{i=-1}^N A_i(s)+Z_N(s), \label{zetad}
\eeq
where, with $\nu=l+\frac{d+1}{2}$ and $m=0$, we have
\beq
A_{-1}(s)= \frac{a^{2s}\Gm(s-\frac{1}{2})}{4\sqrt{\pi}\Gm(s+1)}
     \sma{p} \zeta_H(2s-1-j,\nd),
\eeq
\beq
A_0(s)=-\frac{a^{2s}}{4} \sma{p} \zeta_H(2s-j,\nd),
\eeq
\beq
A_i(s)=-\frac{a^{2s}}{2 \Gm(s)} \sma{p}  \zeta_H(2s+i-j,\nd)
        \sum_{k=0}^i x_{k,i} \frac{(i+2k)\Gm(s+k+\frac{i}{2})}{\Gm(1+k+
        \frac{i}{2})}, \label{aidir}
\eeq
and
\bea
Z_N(s) &=&2sa^{2s} \frac{\sin(\pi s)}{\pi} \sma{p} \sm{l} \nu^{j-2s} \nn \\
&&\times  \itg  dz z^{-2s-1}
  \left\{ \ln I_\nu(\nu z)-\ln\left[ \frac{1}{\sqrt{2\pi \nu}}
 \frac{e^{\nu\mu}}{(1+z^2)^{1/4}}\right]-\sum_{n=1}^N
      \frac{D_n(t)}{\nu^n} \right\}.
\eea
The coefficients $x_{k,i}$ in equation (\ref{aidir}) are obtained from the
polynomial expansion of $D_i(t)$:
\beq
D_i(t)=\sum_{k=0}^i x_{k,i} t^{i+2k}.
\eeq
Similarly, denoting by $z_{i,k}$ the coefficients in
 the expansion
of $M_i(t)$,
\beq
M_i(p,d)(t)=\sum_{k=0}^{2i}z_{i,k}(p,d) t^{i+k},
\eeq
we can write the zeta function for Robin boundary conditions as
\beq
\zeta^{d}_{p}(s)=\sum_{i=-1}^N A_i^R(s)+Z_N^R(s), \label{zetar}
\eeq
where
\beq
A_{-1}^R(s)=A_{-1}(s),
\eeq
\beq
A_0^R(s)=-A_0(s),
\eeq
\beq
A_i^R(s)=-\frac{a^{2s}}{2 \Gm(s)} \sma{p}  \zeta_H(2s+i-j,\nd)
        \sum_{k=0}^{2i} z_{k,i}(p,d) \frac{(i+k)\Gm(s+\frac{i+k}{2})}{\Gm(1+
        \frac{i+k}{2})},
\eeq
and
\bea
Z_N^R(s) &=&2sa^{2s} \left. \frac{\sin(\pi s)}{\pi} \sma{p} \sm{l} \nu^{j-2s}
    \itg  dz z^{-2s-1}
  \right\{ \ln\left[ u(p,d)I_\nu(\nu z)+z\nu I'_\nu(\nu z)\right] \nn \\
&& \left.
    -\ln\left[ \frac{\nu}{\sqrt{2\pi \nu}}
     e^{\nu\mu}(1+z^2)^{1/4}\right]-\sum_{n=1}^N
      \frac{M_n(t)}{\nu^n} \right\}.
\eea

It can be shown that both zeta functions,
(\ref{zetad}) and (\ref{zetar}), are well defined
for $\frac{d-1-N}{2}<s$. Both $Z_N(s)$ and $Z_N^R(s)$ are analytic here,
so that
all the poles are contained in the $A$'s. The analytical continuation can
therefore reach the desired range of $s$, by just changing the value of
$N$.
We are thus in a position where we can find the heat kernel coefficients, the
zeta function determinant and also the Casimir energy.

In the following we shall work in the unit sphere. Since the only change
we have to do on this  zeta function in order
 to include an arbitrary radius is to multiply
by $ a^{2s}$, the results that we obtain for the unit sphere can be
easily converted into those for the general case.
\bs

\section{Calculation of the zeta function at $s=0$}

>From the zeta functions we have defined above, we can now calculate the heat
kernel coefficients, using the relations that exist betwen them.
>From a physical (and maybe also from a  mathematical)
point of view the most interesting coefficient is the one that corresponds
to $\zeta(0)$:
\beq
\zeta(0)=\frac{B_{\frac{d+1}{2}}}{4 \pi^{\frac{d+1}{2}}}, \label{zetaker}
\eeq
where the numerator $B_{\frac{d+1}{2}}$ is the corresponding coefficient
that comes from the short time
expansion of the integrated heat kernel:
\beq
K(t) \sim (4 \pi t)^{\frac{d+1}{2}} \sum_{m=0}^\infty B_{\frac{m}{2}}
      t^{\frac{m}{2}}.
\eeq
Note that $Z_d (0)=0$ and $Z_d^R(0)=0$, since the sum over $l$
and
integral over $z$ is convergent here for $N=d$. Therefore, we need only
consider $A_i(0)$, $i=-1,0,\ldots d$. Using the expansions
\beq
\Gm(-n+\varepsilon)\simeq \frac{(-1)^n}{n!\varepsilon}+..
\eeq
and
\beq
\zeta(1+\varepsilon,\nu)\simeq \frac{1}{\varepsilon}-\Psi(\nu),
\eeq
we  find the expressions
\beq
A_{-1}(0)= \frac{\Gm(-\frac{1}{2})}{4\sqrt{\pi}}
     \sma{p} \zeta_H(-1-j,\nd),
\eeq
\beq
A_0(0)=-\frac{1}{4} \sma{p} \zeta_H(-j,\nd),
\eeq
\beq
A_i(0)=-\frac{1}{4} e_{i-1}
        \sum_{k=0}^i x_{k,i} \frac{(i+2k)\Gm(k+\frac{i}{2})}{\Gm(1+k+
        \frac{i}{2})},
\eeq
and
\beq
A_i^R(0)=-\frac{1}{4}e_{i-1}
        \sum_{k=0}^{2i} z_{k,i}(p,d) \frac{(i+k)\Gm(\frac{i+k}{2})}{
\Gm(1+ \frac{i+k}{2})}.
\eeq
\begin{table}
\begin{center}
\begin{tabular}{|r||c|c|c|c|c|c|c|}                                           									          \hline
$p\backslash d$   &  7                     &  6
&  5                   &  4                   &  3               &  2
     \\ \hline\hline
4               &$-{{3559}\over {9072}}$   &                             &                      &                      &                  &            \\ \hline
3               &${{36583}\over {45360}}$  &${{2929}\over {4608}}$       &${{358}\over {945}}  $&                      &                  &            \\ \hline
2               &$-{{20467}\over {25200}}$ &$-{{1624993}\over {1935360}}$&$-{{1531}\over{1890}}$&$-{{81}\over {128}}$  &$-{7\over {20}}$  &            \\ \hline
1               &${{185449}\over {226800}}$&${{785567}\over {967680}} $  &${{6199}\over {7560}}$&${{2429}\over {2880}}$&${{49}\over {60}}$&${5\over 8}$\\ \hline
0               &$-{{3629089}\over{3628800}}$&$-{{1934993}\over {1935360}}$&$-{{6379}\over {7560}}$&$-{{9647}\over {11520}} $&$-{{151}\over {180}}$ &$-{{41}\over {48}}$ \\ \hline
\end{tabular}
\caption{$\zeta(0)$ for absolute boundary conditions, transversal p-forms.} \label{zeta0abs}
\end{center}
\end{table}

\begin{table}
\begin{center}
\begin{tabular}{|r||c|c|c|c|c|c|c|}                                                                                                                      \hline
$p\backslash d$   &  7                       &  6
&  5                    &  4                  &  3               &  2
     \\ \hline\hline
3                 &$-{{3559}\over {9072}}   $&${{2929}\over {4608}}     $&$                     $&$                   $&$                $&            \\ \hline
2                 &${{4283}\over {25200}}   $&$-{{416417}\over{1935360}}$&${{358}\over {945}}   $&$-{{81}\over {128}} $&$                $&$         $ \\ \hline
1                 &$-{{33521}\over {226800}}$&${{143263}\over {967680}} $&$-{{1109}\over {7560}}$&${{541}\over {2880}}$&$-{7\over {20}}  $&${5\over 8}$\\ \hline
0                 &$-{{289}\over {3628800}} $&$ -{{367}\over {1935360}} $&${1\over {1512}}      $&${{17}\over {11520}}$&$-{1\over {180}} $&$-{1\over {48}}$\\ \hline
\end{tabular}
\caption{$\zeta(0)$ for relative boundary conditions, transversal
p-forms.} \label{zeta0rel} \end{center}
\end{table}
The numerical values obtained from these expressions
 are given in Tables
 \ref{zeta0abs} and \ref{zeta0rel}.
Only those values that are independent have been given (the rest are got
using duality).
For absolute boundary conditions and $p=1$, the scalar field
$B^{0 \bot}$ has a zero mode satisfying the boundary condition
$\partial_0 B^{0 \bot}=0$. But this mode does not contribute in the
Hodge-de Rham decomposition. The scalar field is treated by
construction of the zeta function for Neumann boundary condition. From
the definition of the integrated heat kernel
\beq
K(t)=\sum_n e^{-\lambda_n t},
\eeq
we see that omision the zero mode corresponds to subtracting $1$
from this sum.
Since the relationship (\ref{zetaker}) is still valid ---also when the
zero mode is projected out--- we conclude that in order to get
 $\zeta(0)$ without
zero mode we need to consider $\zeta_{incl}(0)-1$. In this way we
have obtained
the values also for $p=0$, by extracting the values of $\zeta_{incl}(0)$
from~\cite{bek}. For absolute boundary conditions and $d=3$, all values
are in agreement with those calculated in~\cite{Mp}. For relative
boundary
conditions and $d=3$, $p=1$, we have found the value given
in~\cite{vass2}. \bs

\section{Calculation of the determinants}

We will employ the zeta function definition of the determinant of an
operator, $A$, namely
\beq
\ln \det(A)=-\frac{d}{ds} \left.\zeta^A(s)\right|_{s=0}.
\eeq
The determinant of our Laplacian is thus
\beq
-{\ln \det}_{p\perp}^d(\Delta)  = \zeta'^{d}_{pT}  (0)+\zeta'^{d}_{(p-1)T} (0)
\eeq
The determinant for the sphere of radius $a$ is obtained
by adding the terms coming over from the derivation of $a^{2s}$:
\beq
-{\ln \det}_{p\perp}^d(\Delta)(a)  = 2\ln a \left[\zeta^{d}_{pT}  (0)+
           \zeta^{d}_{(p-1)T} (0)\right] + \zeta'^{d}_{pT}  (0)
   +\zeta'^{d}_{(p-1)T} (0).
\eeq
The differentiation of the $A$'s can be done quite easily.
Following the same steps as in Ref.~\cite{bgke}, one obtains the
formulas \bea
Z_d'(0,x)&=&\sma{p} \left\{ \itg dt t^x\frac{e^{-t(\nd)}}{1-e^{-t}} \frac{d^j}{dt^j}
           \left( \frac{t^{-1}}{e^t-1} \right)\right.  \nn\\
&&+\sum_{n=j+1}^d \frac{D_n(1) \Gm(x+n-j)}{\Gm(n-j)} \zeta_H(x+n-j,\nd) \nn\\
&&+\frac{(-1)^j j!}{2} \zeta_H(x-j,\nd)\Gm(x-j) \nn\\
&&\left.-(-1)^j (j+1)!  \zeta_H(x-j-1,\nd)\Gm(x-j-1)\right\}
\eea
and
\bea
Z_d^R ,(0,x)&=&Z_d'(0,x)+\sma{p} \left\{ (-1)^{j+1} \frac{u(p,d)^j+1}{j+1}
              \Psi(\nd) \right. \nn\\
&&+\sum_{n=j+2}^d \frac{u(p,d)^n}{n}(-1)^{n+1} \zeta_H(n-j,\nd) \nn\\
&&-(-1)^j j \int_0^udx x^{j-1} \ln \Gm (\nd+x) \nn \\
&&\left. +(-1)^j u(p,d)\ln \Gm (\nd+u(p,d))\right\}.
\eea
The parameter $x$ is introduced in order to allow for the individual
terms to
 be finite. In the
 final answer this parameter will disappear. The determinants obtained
in this way are listed in the
Appendix. We have also included the determinants for transversal $p=0$
forms given in Ref.~\cite{bgke}. For Neumann boundary
conditions the zero mode must be treated specially, what yelds
an answer we can use directly later on.
\bs

\section{The Casimir energy}

As is well known, the Casimir energy (or vacuum energy) density can be
written as a
 (usually formal)
sum over the eigenvalues of the energy equation, that is
 $\frac{1}{2} \sum_{k} \omega_k$ .
The energy density difference
gives rise to the Casimir force. However, this sum is usually
divergent
and has to be regularized. A very simple and elegant way of
performing the regularization is
to use the zeta function method (see \cite{eorbz,zb2} for very extense
and updated expositions of this procedure). But, sometimes,
it happens that even after analytical continuation the zeta function at the
desired value still diverges. The normal procedure consists then
in resorting to the
principal part prescription~\cite{bvw,ke1}. In Ref. \cite{ke1}
the physical meaning of this prescription has been investigated in depth.
A finite part of the vacuum
energy is found by separating off the pole. Obviosly, from our zeta function,
the vacuum energy is obtained by computing its value at $s=-1/2$.
Writing the zeta function around $s=-1/2$
\beq
\zeta(s)=\frac 1 R \left[ \frac c{s+\frac 1 2} + \phi
+ {\cal O} (s+\frac 1 2)\right],
\eeq
we see that the vacuum energy is given by
\beq
E_C= \frac 1 {2R } \phi.
\eeq
\begin{table}
\begin{center}
\begin{tabular}{|c|c||l|l|} \hline
d & p & absolute boundary conditions&relative boundary conditions  \\
\hline\hline
2 & 0 & $0.008891 + {{2\,\ln (a)}\over {315\,\pi }}$&$0.02806 + {{2\,\ln (a)}\over {45\,\pi }}$\\\cline{2-4}
 & 1 &$0.1678 + {{16\,\ln (a)}\over {315\,\pi }}$&$0.1678 + {{16\,\ln (a)}\over {315\,\pi }}$\\\hline
3&0&$-0.001793$&$-0.03537 - {{2213\,\ln (a)}\over {65536}}$\\\cline{2-4}
 &1&$0.3462 + {{1339\,\ln (a)}\over {32769}}$&$-0.04881 - {{1631\,\ln (a)}\over {65536}} $\\\cline{2-4}
 &2&$-0.04881 - {{1631\,\ln (a)}\over {65536}}$&\\\hline
4&0&$-0.000945 - {{38\,\ln (a)}\over {45045\,\pi }}$&$0.03054 + {{2344\,\ln (a)}\over {15015\,\pi }}$\\\cline{2-4}
 &1&$0.4677 + {{11048\,\ln (a)}\over {45045\,\pi }}$&$0.01881 + {{6632\,\ln (a)}\over {45045\,\pi }}$\\\cline{2-4}
 &2&$ -0.1749 - {{212\,\ln (a)}\over {3465\,\pi }}$&$ -0.1749 - {{212\,\ln (a)}\over {3465\,\pi }}$\\\hline
5&0&$0.0002050$&$-0.02312 - {{3118613\,\ln (a)}\over {50331648}}$\\\cline{2-4}
 &1&$0.5249 + {{871339\,\ln (a)}\over {8388608}}$&$-0.02027 - {{1052991\,\ln (a)}\over {16777216}}$\\\cline{2-4}
 &2&$-0.3459 - {{1063379\,\ln (a)}\over {25165824}}$&$0.05573 + {{31697\,\ln (a)}\over {1048576}}$\\\cline{2-4}
 &3&$0.05573 + {{31697\,\ln (a)}\over {1048576}}$&             \\\hline
\end{tabular}
\caption{$\phi$ for $(d+1)$--dimensional transversal p-forms.}
\end{center}
\end{table}
\begin{table}
\begin{center}
\begin{tabular}{|c||l|l|l|l|l|} \hline
d/p & 5 & 4& 3& 2& 1 \\ \hline\hline
2   & & & & 0.1959&0.1767\\\hline
3   & & &$-0.08417$&0.2974&0.3444 \\\hline
4   &   &  0.04935&$-0.1561$&0.2928&0.4668  \\\hline
5   &$-0.04340 $&$0.03546 $&$-0.2902 $&$0.1790$&$0.5251 $   \\\hline
\end{tabular}
\caption{$\phi$ for  p-forms on the unit sphere with absolute boundary
 conditions.}
\end{center}
\end{table}
At $s=-1/2$ one observes that the poles come from the gamma function of
$A_{-1}$, from the gamma function of $A_i$, for $b=0$ and $i=1$, and from
 the zeta function of $A_i$, when $i=m+2$. We perform a Laurent expansion
around these poles and isolate the corresponding finite parts.
The rest of the functions will only contribute to the finite part.
The values
$Z_{d+1}(-\frac{1}{2})$ and $Z_{d+1}^R(-\frac{1}{2})$ have to be computed
numerically. These contributions are generally quite small compared with
the finite part which comes from $\sum_{i=-1}^{N} A_i
$. By adjusting $N$,
the values of
$Z_{d+1}^{(R)}(-\frac{1}{2})$ can be further reduced, allowing one to
obtain the same accuracy with much less effort. For some values of $d$
and $p$ the sum over $l$ converges very slowly. Use of Richardson
extrapolation leads to a dramatical improvement of the convergence
speed.
This extrapolation is a general procedure of numerical analysis. It
is here valid because the partial sum has the following asymptotic
behavior: \beq
\sum_{l=0}^n \nu^{j+1}\int_{0}^\infty \left\{ \ldots \right\} \sim
   Q_0 +Q_1 n^{-1}+Q_2 n^{-2}+Q_3 n^{-3}+ \ldots , n \rightarrow \infty.
\eeq
The finite contributions for the $(d+1)$--transversal forms are listed
in Table 3.
We have included the scalar field, $p=0$. When the argument of the zeta
function is negative, the constant term for Neumann boundary conditions
does not contribute. Special care of this term need
therefore not be taken here. The coefficients belonging to $\ln (a)$
 equal  the
heat-kernel coefficient
$-\frac{B_{\frac{d}{2}+1}}{(4 \pi)^{\frac{d+1}{2}}\sqrt{\pi}}$.
Table 4 gives the vacuum energy for the unit sphere for all
$p$-forms, for absolute boundary condition.
For $p=1$ we see that the energy
increases with increasing $d$. For $p=2$ there is a maximum
 at $d=3$, while
the energy for $p=3$ and $p=4$ decreases with $d$. For constant
$d$ there are
actually less systematic trends.

\vspace{4mm}

\noindent{\bf Acknowledgments}

We are grateful to Klaus Kirsten and Sergio Leseduarte for
very interesting discussions. ML would like to thank the members
of the Department ECM, University of Barcelona, for their warm
hospitality during.
The stay of ML in Barcelona was made possible by the ERASMUS program.
EE was supported by DGICYT, project PB93-0035,
 and by the Ministry of Foreign Affairs
(Spain). DV was supported by the Russian
Foundation for Fundamental Research and by GRACENAS, project
M94-2.2$\Pi$-18.
\bs

\appendix
\section{The zeta function determinants}
\subsection{Absolute boundary conditions}

In this case, we have obtained
\bea
-{\ln \det}_{3\perp}^5(\Delta)&=& { {55120073}\over {64864800}} -
   {{4\, }\over 3} \int _{0}^{1}dy\,{y^2}\,{\rm \Psi}(3 + y)
  + {{1}\over 3}\int _{0}^{1} dy\,{y^4}\,{\rm \Psi}(3 + y) \nn\\
&& -
   {{29\,\ln 2}\over {3780}}- {{215\,\ln 3}\over 2} +
   {{\ln 4}\over 8} + {{{\rm \zeta_R}'(-5)}\over 6} -
   {{{\rm \zeta_R}'(-4)}\over {12}}\nn\\
&&   - {\rm \zeta_R}'(-3) +
   {{7\,{\rm \zeta_R}'(-2)}\over {12}} + {{4\,{\rm \zeta_R}'(-1)}\over 3} -
   {\rm \zeta_R}'(0)  \nn\\
-{\ln \det}_{2\perp}^5(\Delta)&=& {{ -38814043}\over {64864800}} +
   {{5179\,\ln 2}\over {3780}}  -
   {{3\,\ln 4}\over 8} + {{{\rm \zeta_R}'(-5)}\over 6}\nn\\
&& +
   {{{\rm \zeta_R}'(-4)}\over {12}} - {\rm \zeta_R}'(-3) -
   {{7\,{\rm \zeta_R}'(-2)}\over {12}} + {{4\,{\rm \zeta_R}'(-1)}\over 3} +
   {\rm \zeta_R}'(0)\nn\\
-{\ln \det}_{1\perp}^5(\Delta)&=& {{ 75711793}\over {64864800 }} +
   {{8\, }\over 3} \int _{0}^{-1}dy\,y\,\ln {\rm \Gm}(3 + y)
    - {{4\, }\over 3} \int _{0}^{-1}  dy\,{y^3}\,\ln {\rm \Gm}(3 + y)\nn\\&&
  -  {{3\,\ln ({1\over 2})}\over 2} - {{11869\,\ln 2}\over {3780}} -
   + {{{\rm \zeta_R}'(-5)}\over {12}} +
   {{{\rm \zeta_R}'(-4)}\over 8} \nn\\&&- {{{\rm \zeta_R}'(-3)}\over 3} -
   {{5\,{\rm \zeta_R}'(-2)}\over 8} - {{{\rm \zeta_R}'(-1)}\over 4} \nn\\
-{\ln \det}_{0\perp}^5(\Delta)&=&-{{7087979}\over {32432400}} - {{1181\,\ln 2}\over {3780}} +
   \ln 3 +{{{\zeta_R}'(-5)}\over {60}} + {{{\zeta_R}'(-4)}\over {24}} -
   {{{\zeta_R}'(-2)}\over {24}} - {{{\zeta_R}'(-1)}\over {60}}\nn\\
& &-\frac 1 6 \int\limits_0^2dy\,\,(y-2)\ln\Gm (y) +\frac 1 3
\int\limits_0^2 dy\,\,(y-2)^3\ln\Gm (y)\nn\\
-{\ln \det}_{2\perp}^4(\Delta)&=& -{{3411}\over {2560}} +
   {{9\,  }\over 4}\int _{0}^{{1\over 2}}  dy\,y\,{\rm \Psi}({5\over 2} + y) -
   \int _{0}^{{1\over 2}}dy\,{y^3}\,{\rm \Psi}({5\over 2} + y) -
   {{45\,\ln 2}\over {64}}  \nn\\&& - {{15\,{\rm \zeta_R}'(-4)}\over {32}} +
   {{21\,{\rm \zeta_R}'(-2)}\over {16}}\nn\\
-{\ln \det}_{1\perp}^4(\Delta)&=& {{10633}\over {11520}} -
   {{9\,  }\over 4} \int _{0}^{-{1\over 2}}dy\,\ln {\rm \Gm}({5\over 2} + y)
  + 3\, \int _{0}^{-{1\over 2}}  dy\,{y^2}\,
\ln {\rm \Gm}({5\over 2} + y)  +
   \ln ({2\over 3}) \nn\\&& + {{581\,\ln 2}\over {2880}} -
   {{5\,{\rm \zeta_R}'(-4)}\over {16}} - {{7\,{\rm \zeta_R}'(-3)}\over {24}} +
   {{5\,{\rm \zeta_R}'(-2)}\over 8} + {{13\,{\rm \zeta_R}'(-1)}\over {24}}\nn\\
-{\ln \det}_{0\perp}^4(\Delta)&=&-{{19261}\over {46080}} - {{713\,\ln 2}\over {72   0}} + \ln 5 - {{5\,{\zeta_R}'(-4)}\over {64}} -
   {{7\,{\zeta_R}'(-3)}\over {48}} - {{{\zeta_R}'(-2)}\over {32}} +
   {{{\zeta_R}'(-1)}\over {48}}\nn\\
& &+\frac 1 {12} \int\limits_0^{3/2}dy\,\,\ln\Gm (y) -
    \int\limits_0^{3/2} dy\,\,\left(y-\frac 3 2 \right)^2 \ln\Gm (y)\nn\\
-{\ln \det}_{2\perp}^3(\Delta)&=& -{{2081}\over {3360}} +
    \int _{0}^{1}dy\,{y^2}\,{\rm \Psi}(2 + y) -
    {{\ln 2}\over {20}} - {{\ln 4}\over 8} + {\rm \zeta_R}'(-3) -
    {{{\rm \zeta_R}'(-2)}\over 2}\nn\\
& & - {{3\,{\rm \zeta_R}'(-1)}\over 2} + {\rm \zeta_R}'(0)\nn\\
-{\ln \det}_{1\perp}^3(\Delta)&=& {{5989}\over {10080}} - {{83\,\ln 2}\over {60}} +
   {{3\,\ln 4}\over 8} + {\rm \zeta_R}'(-3) + {{{\rm \zeta_R}'(-2)}\over 2} -
   {{3\,{\rm \zeta_R}'(-1)}\over 2} - {\rm \zeta_R}'(0)\nn\\
-{\ln \det}_{0\perp}^3(\Delta)&=&-{{493}\over {4320}} + {{61\,\ln 2}\over {90}}
 + {{\zeta_R'(-3)}\over 3} +
   {{\zeta_R'(-2)}\over 2} + {{\zeta_R'(-1)}\over 6}
+2\int\limits_0^1dy\,\,(y-1)\ln\Gm (y)\nn\\
-{\ln \det}_{1\perp}^2(\Delta)&=& \frac{19}{16} + 2\, \int _{0}^{{1\over 2}}
          dy\,\ln {\rm \Gm}({3\over 2} + y)   +\frac 3 4\,\ln 2 -
     \frac 3 2\,{\rm \zeta_R}'(-2) \nn\\
-{\ln \det}_{0\perp}^2(\Delta)&=& -{7\over {32}} - {{7\,\ln 2}\over 6} + \ln 3 - {{3\,{\zeta_R}'(-2)}\over 4} -
   {{{\zeta_R}'(-1)}\over 2}-2\int\limits_0^{1/2}dy\,\,\ln\Gm (y)
\eea

\subsection{Relative boundary conditions}

In this case, the results are
\bea
-{\ln \det}_{2\perp}^5(\Delta)&=& {{55120073}\over {64864800}} -
   {{4\,   }\over 3}\int _{0}^{1}dy\,{y^2}\,{\rm \Psi}(3 + y)
  + {{1}\over 3}\int _{0}^{1}  dy\,{y^4}\,{\rm \Psi}(3 + y) \nn\\
& &   - {{29\,\ln 2}\over {3780}}
 - {{215\,\ln 3}\over 2} +
   {{\ln 4}\over 8} + {{{\rm \zeta_R}'(-5)}\over 6} -
   {{{\rm \zeta_R}'(-4)}\over {12}} - {\rm \zeta_R}'(-3) \nn\\
&& +  {{7\,{\rm \zeta_R}'(-2)}\over {12}} + {{4\,{\rm \zeta_R}'(-1)}\over 3} -
   {\rm \zeta_R}'(0)\nn\\
-{\ln \det}_{1\perp}^5(\Delta)&=& {{ -11173163}\over {64864800}} -
   {{1}\over {12}}\int _{0}^{2}dy\,{y^2}\,{\rm \Psi}(3 + y) +
   {{1}\over {12}}\int _{0}^{2}dy\,{y^4}\,{\rm \Psi}(3 + y) -
   {{3\,\ln ({1\over 2})}\over 2} \nn\\
& &- {{8341\,\ln 2}\over {3780}}
   + {{{\rm \zeta_R}'(-5)}\over {12}} -
   {{{\rm \zeta_R}'(-4)}\over 8} - {{{\rm \zeta_R}'(-3)}\over 3} \nn\\
& & + {{5\,{\rm \zeta_R}'(-2)}\over 8} - {{{\rm \zeta_R}'(-1)}\over 4}\nn\\
-{\ln \det}_{0\perp}^5(\Delta)&=& -\frac{4027}{6486480} -\frac 1 {756}\ln 2
    +\frac  1 {60}\,\zeta_R'(-5) -\frac 1 {24}\,\zeta_R'(-4)
    +\frac 1 {24}\,\zeta_R'(-2) -\frac 1 {60}\,\zeta_R'(-1)\nn\\
-{\ln \det}_{2\perp}^4(\Delta)&=& -{{3411}\over {2560}} +
   {{9\,}\over 4} \int _{0}^{{1\over 2}} dy\,y\,{\rm \Psi}({5\over 2} + y) -
   \int _{0}^{{1\over 2}}dy\,{y^3}\,{\rm \Psi}({5\over 2} + y) -
   {{45\,\ln 2}\over {64}}  \nn\\
& & - {{15\,{\rm \zeta_R}'(-4)}\over {32}} +
   {{21\,{\rm \zeta_R}'(-2)}\over {16}}\nn\\
-{\ln \det}_{1\perp}^4(\Delta)&=& {{17021}\over {11520}} +
   {{1}\over {12}}\int _{0}^{{3\over 2}}dy\,y\,{\rm \Psi}({5\over 2} + y)
 - {{1}\over 3}\int _{0}^{{3\over 2}} dy\,{y^3}\,{\rm \Psi}({5\over 2} + y) -
   {{2561\,\ln 2}\over {2880}}\nn\\
&&- {{5\,{\rm \zeta_R}'(-4)}\over {16}}+
   {{7\,{\rm \zeta_R}'(-3)}\over {24}} + {{5\,{\rm \zeta_R}'(-2)}\over 8} -
   {{13\,{\rm \zeta_R}'(-1)}\over {24}}\nn\\
   -{\ln \det}_{0\perp}^4(\Delta)&=&\frac{47}{9216} + \frac{17}{2880}\ln 2
    +\frac 5 {64}\,\zeta_R'(-4)+\frac 7 {48}\,\zeta_R'(-3) -
   \frac 1 {32}\,\zeta_R'(-2) - \frac 1 {48}\,\zeta_R'(-1)\nn\\
-{\ln \det}_{1\perp}^3(\Delta)&=&- {{2081}\over {3360}} +
   \int _{0}^{1}dy\,{y^2}\,{\rm \Psi}(2 + y) -
   {{\ln 2}\over {20}} - {{\ln 4}\over 8} + {\rm \zeta_R}'(-3)\nn\\
&& -    {{{\rm \zeta_R}'(-2)}\over 2}
    - {{3\,{\rm \zeta_R}'(-1)}\over 2} + {\rm \zeta_R}'(0)\nn\\
-{\ln \det}_{0\perp}^3(\Delta)&=&\frac{173}{30240} +\frac 1 {90} \ln 2 + \frac 1    3 \zeta_R'(-3)\nn\\
-{\ln \det}_{1\perp}^2(\Delta)&=& \frac{19}{16} + 2\, \int _{0}^{{1\over 2}}
          dy\,\ln {\rm \Gm}({3\over 2} + y)   +\frac 3 4\,\ln 2 -
     \frac 3 2\,{\rm \zeta_R}'(-2) \nn\\
-{\ln \det}_{0\perp}^2(\Delta)&=&-\frac 3 {32} -\frac 1 {12}\ln 2 -\frac 3 4
\zeta_R'(-2)+\frac 1 2 \zeta_R'(-1)
\eea

\end{document}